\documentclass[showpacs,aps,pre,twocolumn]{revtex4}

\usepackage{graphicx}
\usepackage{dcolumn}
\usepackage{bm}
\usepackage{url}

\usepackage{amssymb}
\usepackage{amsmath}

\newcommand{\eps}{\epsilon}

\makeatletter

\begin{document}

\title{Statistical mechanics of the lattice sphere packing problem}

\author{Yoav Kallus}
\email{ykallus@princeton.edu}
\affiliation{Princeton Center for Theoretical Science, Princeton University, Princeton, New Jersey 08544}

\date{\today}

\begin{abstract}
We present an efficient Monte Carlo method for the lattice sphere packing problem
in $d$ dimensions. We use this method to numerically discover \textit{de novo}
the densest lattice sphere packing in dimensions 9 through 20. Our method
goes beyond previous methods, not only in exploring higher dimensions but also in
shedding light on the statistical mechanics underlying the problem in question.
We observe evidence of a phase transition in the thermodynamic limit $d\to\infty$. In the
dimensions explored in the present work, the results are consistent with
a first-order crystallization transition but leave open the possibility
that a glass transition is manifested in higher dimensions.
\end{abstract}
\pacs{61.50.Ah, 05.10.Ln, 05.20.Jj}

\maketitle

The problem of identifying the highest density sphere packing in $d$ dimensions
is a classical problem in geometry with direct connections to fields of physics,
information theory, and mathematics. The case $d=3$, for which Kepler conjectured
that the face-centered cubic lattice is optimal, stood as an open problem for
centuries before, finally, in 1998, a proof was announced by Hales \cite{HalesKepler}.
Aside from dimensions $2$ and $3$, the highest density is not known in any
other dimension, although it has been bounded to an extremely tight interval
in dimensions $8$ and $24$ \cite{CohnKumar}.

The highest packing densities that have been obtained in these dimensions, as in many others, are obtained
by Bravais lattices: periodic packings with one sphere in each unit cell. From this point on,
we use ``lattice'' to mean a Bravais lattice unless a non-trivial basis is mentioned.
When restricting the sphere packing problem to lattices, many simplifications are possible,
and the problem becomes more tractable, but still far from trivial. In fact, the densest
lattice packings are known in dimensions $d\le 8$ and $d=24$ \cite{SPLAG,CohnKumar}. The space of lattice packings
in $d$ dimensions is finite-dimensional and much simpler to study than the infinite-dimensional
space of all possible packings.

In fact, it is possible, in principle, to identify all local density maxima in this space.
Such lattices, called extreme lattices, have been characterized by Voronoi in terms of their algebraic
properties. Voronoi showed that a lattice is extreme if and only if it is perfect
and eutactic \cite{perfect}. Perfect lattices---those that are
fully determined by a list of their shortest vectors---are finite in number,
and Voronoi gave an algorithm that enumerates all the perfect lattices in
a given dimension. In dimensions $d\le 8$, the identity of the densest lattice
packing has been established by an exhaustive enumeration of the perfect lattices.
Voronoi's algorithm relies on a method of obtaining, starting from any
perfect lattice, a set of \textit{neighboring} perfect lattices. Voronoi showed that
this graph of neighbors is connected, so exploring larger and larger neighborhoods
of a single perfect lattice would eventually uncover all perfect lattices.

However, as the number of perfect lattices grows rapidly in dimensions $d>8$,
exhaustive enumeration becomes impractical, and other methods must be used to identify
dense packings. Analytic constructions based on groups, codes, and laminated lattices
have been successful in producing very dense lattice
sphere packings in dimensions up to $d=24$ and in certain
dimensions above \cite{SPLAG}.
While these methods have certainly proved remarkably effective, they
give little reason, on their own, to believe that they have, in fact,
produced the densest possible structures.

For the latter purpose, in the absence of rigorous proofs (as in $d=24$)
or tight bounds, we rely on numerical methods that attempt to discover
these structures \textit{de novo}: without \textit{a priori} knowledge of their
existence. It is only recently that such methods were introduced that
could tackle moderately high dimensions. A method based on the ``divide
and concur'' framework for constraint satisfaction problems was
used in Ref. \cite{kallusDC} to discover \textit{de novo} the densest known
lattices in $d\le14$. In Ref. \cite{random-perfect}, Andreanov and Scardicchio
implemented a random walk on Voronoi's graph, yielding
a random sample of perfect lattices. While their method was designed
to explore random perfect lattices and their statistics, they were also
able to use it for \textit{de novo} discovery of the densest lattice packing.
In dimensions $8\le d\le12$,
their random sample included the densest known lattice packing, while in
dimensions $13\le d\le 19$, they had to bias their random steps
toward higher density configuration to recover the densest known packing.
In $d=19$, only some of these biased walks ended up visiting the densest
known lattice.
In Ref. \cite{marcotte}, Marcotte and Torquato used a sequential linear programming (SLP)
approach to iteratively compress a lattice configuration until it reaches
a configuration that cannot be compressed any more---an extreme lattice.
This procedure, starting from random initial conditions, reproduced the
densest known lattice in an appreciable portion of the runs in each
dimension $d\le16$. The percentage of runs yielding the densest known lattice declines
sharply starting at $d=17$. However, because each run can still be computed
rapidly, the procedure can be repeated many times and the densest known
lattice is produced at a decent rate for $d\le19$ (see below for a direct
comparison with the present method). Other methods
have been used for \textit{de novo} searches in closely related problems,
such as the Gaussian-core soft sphere ground state problem \cite{gauss-core} and
the lattice quantizer problem \cite{quantizer}.

In this paper we report on a Monte Carlo (MC) method for studying the
lattice sphere packing problem. Our method is completely different
from the references above, but some of our results are surprisingly similar
to the results of Refs. \cite{random-perfect,marcotte}.
In particular, in each dimension $d\le19$, a simulated quasistatic compression discovers
\textit{de novo} the densest known lattice in at least $30\%$ of the runs.
In 20 dimensions, only 1 of 50 quasistatic compression runs
yielded the densest known lattice. However, with a slight change
in protocol, we were able to reproduce the lattice in 7 of 50
runs. That three such disparate methods all seem to start having
serious difficulties in the same dimensions raises
the possibility that the lattice sphere packing
problem becomes intrinsically harder around $d=20$. We suggest
possible reasons for such a scenario.
Moreover, because of the statistical mechanical nature of our
method, we are able to quantitatively characterize the 
intrinsic nature of the lattice sphere packing problem apart
from the behavior of any specific
method or algorithm for its solution.

A lattice can be specified by its generating matrix: the set of all sphere centers is given
by $M^T\mathbb Z^d=\lbrace M^T\mathbf{n}:\mathbf{n}\in\mathbb{Z}^d\rbrace$,
where $M$ is a $d\times d$ matrix and its rows are generating vectors
(primitive vectors) of the lattice. While a generating matrix uniquely specifies
a lattice, a lattice has multiple generating matrices: whenever $Q$ is a unimodular
integer matrix,  $M'=QM$ generates the same lattice as $M$.
A lattice is a packing of radius-$1/2$ spheres (that is, its spheres do not overlap) if
$||M^T\mathbf{n}||^2\ge1$ for all $\mathbf{n}\in\mathbb{Z}^d\setminus\lbrace0\rbrace$.
If this is the case, we say the lattice is \textit{admissible}.
The number density of spheres is given by $1/v=1/|\det M|$,
where $v$ denotes the unit cell volume. Any lattice, generated by $M$,
can be rotated so that its generating matrix $MU$, where $U$ is a rotation matrix, becomes
lower-triangular. Therefore, the space of lattices, modulo rotation, can be parameterized
by lower-triangular generating matrices.

We can define an isobaric ensemble on the space of admissible lattices by weighting the
probability of each lattice by a factor $\exp(-p v)$, where $p=NP/k_B T$ is
a reduced pressure variable (cf. Ref. \cite{parisi} for definitions
of related ensembles). If $v$ is thought of as the energy
associated with a particular lattice, we can think of $p$ as the inverse-temperature
(cf. Ref. \cite{random-perfect}, where $\log v$ is used as the energy).
We can sample this ensemble using a
standard Metropolis algorithm: a random element of the lower-triangular
generating matrix is randomly changed by a small amount;
the step is rejected always if it yields an inadmissible lattice
and is rejected with probability $\exp(-p\Delta v)$ if the
unit cell volume is increased by $\Delta v$. However,
we find instead that it is more efficient to build the detailed balance
directly into the proposed steps instead of the acceptance probability.
Note that changes to the off-diagonal elements of the matrix do not change
the volume. Therefore, these moves are always accepted if they produce
admissible lattices. For the diagonal elements, we propose changes
\begin{equation*}
m_{ii}\leftarrow\left(1+\frac{\eps x-\tfrac{1}{2}\eps^2p}{v}\right) m_{ii}\text,
\end{equation*}
where $v=\prod_{j=1}^d m_{jj}$ is the current volume, $\eps$ is a measure
of the typical move size, and $x$ is drawn from a normal distribution of unit variance.
As with the off-diagonal moves, the proposed moves are always accepted
if they produce admissible lattices. 
Note that the change in volume is given by
$\Delta v = \Delta m_{ii} v/m_{ii} = \eps x - \tfrac{1}{2}\eps^2 p$
and is distributed with a probability density of
\begin{equation*}
\frac{1}{\eps\sqrt{2\pi}} \exp\left(-\frac{(\Delta v + \tfrac{1}{2}\eps^2 p)^2}{2\eps^2}\right)\text.
\end{equation*}
Therefore, an accepted move
changing the volume by $\Delta v>0$ is less likely by a factor of
$\exp(p\Delta v)$ than the reverse move, as required by detailed balance.

A crucial step in this MC algorithm is checking whether a lattice
is admissible. This is known to be an NP-complete problem \cite{lattpt-comp},
and in fact it takes up most of the computational time in our simulations.
The complexity of this problem is sensitive to the choice of
generating matrix for a given lattice. Generally speaking, the shorter and more
orthogonal to each other the generating vectors are, the easier the
problem is of determining admissibility of the lattice they generate.
There are many non-equivalent criteria for determining how well-suited,
or \textit{reduced}, a certain set of generating vectors is. We use
Korkine-Zolotareff (KZ) reduction, which is one of the most stringent of
these criteria \cite{agrell}. Such
a stringent criterion is warranted because of the large number of times
we are required to decide the admissibility of similar lattices. Therefore,
during our simulation we periodically perform KZ reduction, and thus all
the generating matrices we consider are either KZ-reduced or nearly so.
Detailed balance is violated by these reductions, but we find that
the reductions are rare enough in the key stages of the simulations
that they do not significantly hinder thermalization.

Using our MC technique, we perform a simulated
quasistatic compression (a simulated annealing where pressure takes
the role of temperature). We start the system in a simple hypercubic
lattice $\mathbb{Z}^d$ and equilibrate at a constant pressure.
We then start to increase the pressure by a constant factor after
each proposed move. We vary the typical move size inversely with the
pressure: $\eps=\eps_0/p$. We use different move sizes
for off-diagonal and diagonal moves, and for both we pick
$\eps_0$ so as to achieve an average move acceptance rate
of roughly $30\%$. The length of the equilibration period is $4\%$
of the length of the compression period.
The Gram matrix $G=M M^T$ of the final matrix obtained in nearly all
runs consists, up to small errors, of small-denominator
rational numbers. Therefore, we can easily round off its
entries to obtain the infinite pressure limit of the simulation.

\begin{table}\begin{center}
\begin{tabular}{|c|c|c|c|c|c|c|c|}
\hline
$d$ & $\Lambda$ & $2^d v$ & $p_i$ & $p_f$ &
   $k$ & $T$ (sec.) & rate \\\hline
$9$ & $\Lambda_9$ & $16\sqrt{2}$ & $20$ & $2\times10^4$ &
   $1.4\times10^{-6}$ & $60$ & $1$ \\
$10$ & $\Lambda_{10}$ & $16\sqrt{3}$ & $60$ & $6\times10^4$ & 
   $1.4\times10^{-6}$ & $1.2\times10^2$ & $1$ \\
$11$ & $K_{11}$ & $18\sqrt{3}$ & $2\times10^2$ & $2\times10^5$ &
   $4.6\times10^{-7}$ & $7.8\times10^2$ & $0.55$ \\
$12$ & $K_{12}$ & $27$ & $3\times10^2$ & $3\times10^5$ &
   $6.9\times10^{-7}$ & $1.1\times10^3$ & $1$  \\
$13$ & $K_{13}$ & $18\sqrt{3}$ & $6\times10^2$ & $6\times10^5$ &
   $3.5\times10^{-7}$ & $3.0\times10^3$ & $0.70$ \\
$14$ & $\Lambda_{14}$ & $16\sqrt{3}$ & $2\times10^3$ & $2\times10^6$ &
   $1.8\times10^{-7}$ & $9.4\times10^3$ & $0.60$ \\
$15$ & $\Lambda_{15}$ & $16\sqrt{2}$ & $3\times10^3$ & $3\times10^6$ &
   $3.5\times10^{-7}$ & $9.1\times10^3$ & $0.90$ \\
$16$ & $\Lambda_{16}$ & $16$ & $5\times10^3$ & $5\times10^6$ &
   $1.8\times10^{-7}$ & $2.9\times10^4$ & $0.95$ \\
$17$ & $\Lambda_{17}$ & $16$ & $1.5\times10^4$ & $1.5\times10^7$ &
   $6.9\times10^{-8}$ & $1.2\times10^5$ & $0.8$ \\
$18$ & $\Lambda_{18}$ & $8\sqrt{3}$ & $7\times10^4$ & $1.8\times10^7$ &
   $5.5\times10^{-8}$ & $1.9\times10^5$ & $0.6$ \\
$19$ & $\Lambda_{19}$ & $8\sqrt{2}$ & $9\times10^4$ & $2.2\times10^7$ &
   $4.6\times10^{-8}$ & $2.4\times10^5$ & $0.3$ \\
$20$ & $\Lambda_{20}$ & $8$ & $3\times10^5$ & $8.2\times10^6$ &
   $3.5\times10^{-8}$ & $3.3\times10^5$ & $0.14$ \\\hline
\end{tabular}
\caption{\label{tabres} For each dimension $d$, the table gives the following:
the name (as per Refs. \cite{SPLAG,nebe}) of the lattice $\Lambda$
that achieves the greatest density known among all admissible lattices;
the unit cell volume $v$ of this lattice (normalized by $2^d$); the
reduced pressure $p_i$ used in the equilibration period and at the
beginning of the compression period of our simulations; the
reduced pressure $p_f$ at which we terminated the compression;
the compression rate $k$, such that the pressure at each proposed move
is $1+k$ times the pressure at the previous proposed move;
the average computational time $T$ per run of the simulation; and the
rate at which the lattice $\Lambda$ is reproduced, that is, the percentage
of runs whose final configuration, after rounding off the Gram matrix, is $\Lambda$.}
\end{center}\end{table}

For each dimension $d=9,\ldots,19$ we performed $20$ simulation runs.
In each of these dimensions, our simulations discover the densest
known lattice packing in at least $30\%$ of the runs. Table \ref{tabres} 
summarizes the results of these simulations.
The compression rate used
represents a trade-off between longer computation time
and decreased likelihood of reproducing the densest lattice.
We did not attempt to quantify this trade-off in this paper or determine
the optimal compression rate as a function of dimension.
In terms of the average computational time needed
to reproduce the densest known lattice once, the present method is much less efficient
than the SLP method of Ref. \cite{marcotte} in all lower dimensions.
For the highest dimensions, $d=17,18,$ and $19$, respectively, this average time is
$3\times10^3 \mathrm{s},$ $8 \times 10^4\mathrm{s}$, and $2\times 10^6\mathrm{s}$
with the SLP method, compared to $2\times 10^5\mathrm{s}$, $3\times 10^5\mathrm{s}$,
and $8\times 10^5 \mathrm{s}$ with the present method.

\begin{figure}\begin{center}
\includegraphics[scale=0.6]{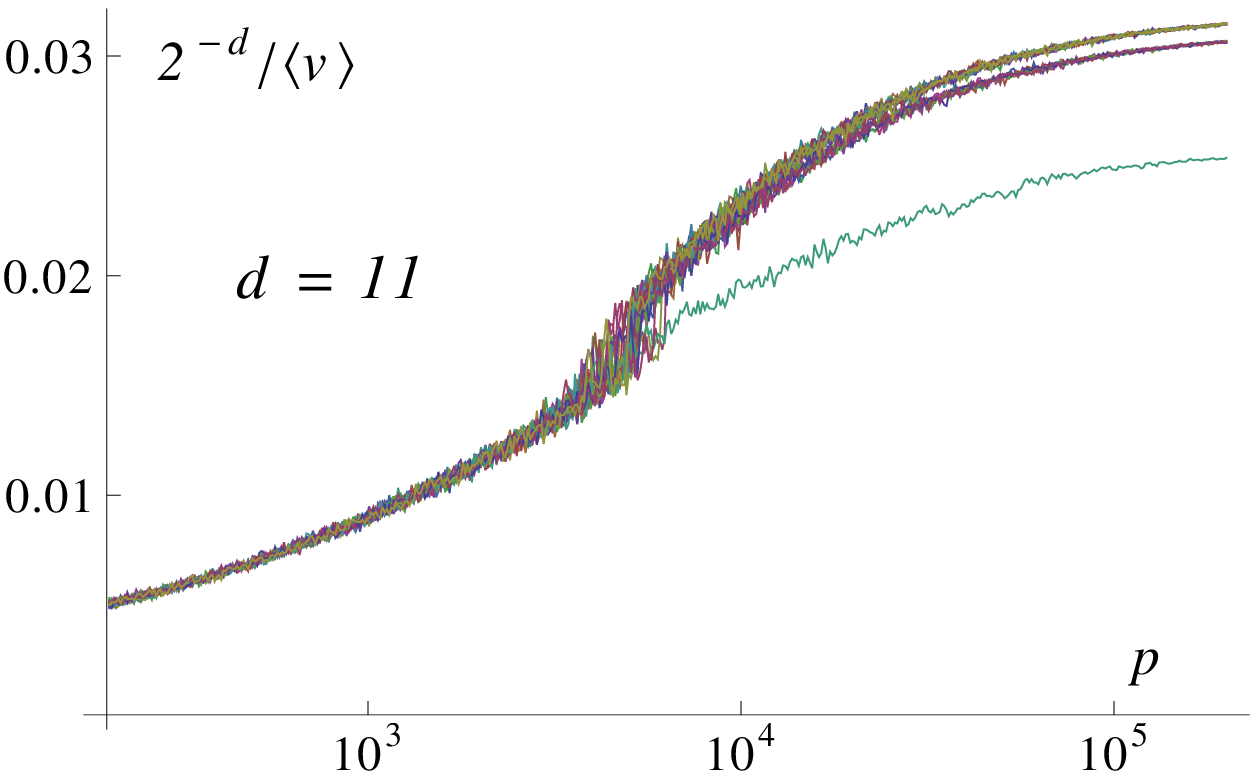}
\includegraphics[scale=0.6]{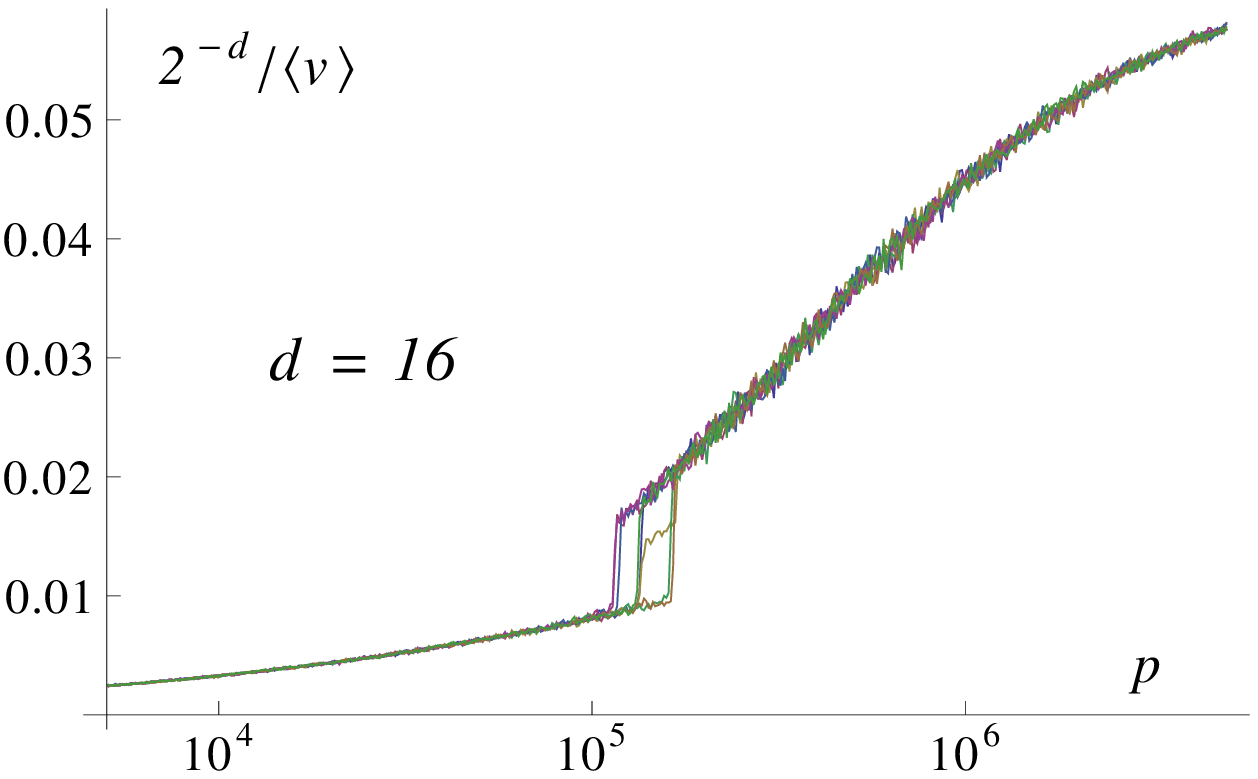}
\caption{\label{figd11}\label{figd16}Traces of density as a function of reduced
pressure in different runs in dimensions $d=11$ and $d=16$. In 11 dimensions,
the simulation remains in equilibrium until around $p=4.8\times10^3$,
where different runs continue along different branches, corresponding to
basins of attractions of different extreme lattices. In 16 dimensions,
all runs end up in the basin of the same lattice, but we observe that
different runs make the transition into this
basin at different pressures. It appears that at least one of the runs
also spends some time in an intermediate state, presumably the
basin of attraction of a different extreme lattice.}
\end{center}\end{figure}

\begin{figure}\begin{center}
\includegraphics[scale=0.6]{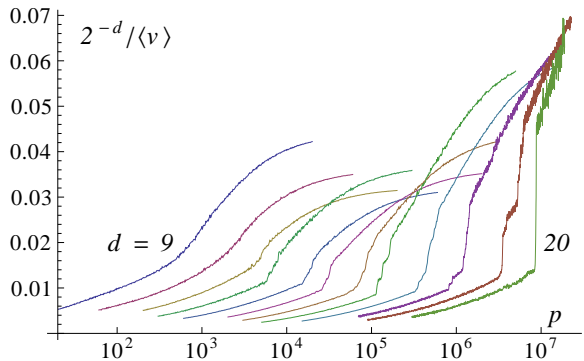}
\caption{\label{figall}Traces of (normalized) density as a function of reduced
pressure, averaged over all quasistatic compression runs that yielded the densest
known lattice in each dimension $d=9,\ldots,20$.}
\end{center}\end{figure}

Assuming our MC simulations accurately sampled
the isobaric ensemble at each pressure, we would
recover from the simulation the equation of state: the
average volume $\langle v\rangle$ as a function of the reduced pressure $p$.
Plotting the traces of $\langle v\rangle$ as a function
of $p$ for different runs in dimension $d=11$ for example (Fig. \ref{figd11}),
we see that at a certain pressure the different traces
diverge and the simulations fall
out of equilibrium. The traces belonging to runs that terminate
at the same configuration do not diverge, so we may infer that
at this pressure the system goes from exploring
the attraction basins of many different extreme lattices
to exploring only a single basin. The situation becomes
more complicated in higher dimensions, where we see significant
hysteresis effects. For example, in dimension $d=16$, all 20
runs end up in the basin of the densest known lattice $\Lambda_{16}$ (Fig. \ref{figd16}),
but different runs experience the transition at different pressures.
Figure \ref{figall} shows, for each dimension, the volume
as a function of pressure averaged over all the runs that yielded
the densest known packing.

For any fixed $d$, as there is no
thermodynamic limit, strictly speaking, there
cannot be a phase transition. However, as
is clear from Figs. \ref{figd11} and \ref{figall},
the system shifts as the pressure increases from a state
where many basins of attraction are explored
to a state where the system is confined to a single basin.
In any finite dimension there should be a range of pressure
where these two states coexist with significant probability
for the system to be in either state.
The traces we obtain from the simulations are consistent
with a situation where the transition rate between these
two states in the coexistence region becomes smaller and
smaller with increased dimension, so that in lower dimensions
the trace of each run remains close to the equation of state,
while in higher dimensions each run stays in one state
until transitioning irreversibly into the other.
This crystallization transition is accompanied
by a discontinuity in the density, which
depends on the extreme lattice the system
crystallizes into.

We may interpret $d\to\infty$ as the thermodynamic limit of
the lattice sphere packing problem and speculate that
the coexistence region in this limit shrinks to a single 
critical pressure. Parisi considers the partition
function of a closely related ensemble of lattices in
the limit $d\to\infty$ and points out
that the lattice sphere packing problem in this limit
shares many formal similarities with the nonlattice problem
\cite{parisi}. Parisi does not determine whether
a glass transition exists in the lattice sphere packing problem
as it does in nonlattice hard sphere system,
and leaves open the possibility of either a glass transition
or a crystallization transition.
The behavior we observe in our simulations
in dimensions $9\le d\le 19$ is indicative of a first-order
crystallization transition. However, it is hard to tell whether
the behavior is controlled by the thermodynamic limit
or mostly by details specific to the moderate dimensions
we explore.

In dimension $d=20$, out of 50 runs at the slowest compression rate
attempted, only one yielded the
densest known lattice. In fact, only ten runs show
a discontinuity in the density at all, with most
runs remaining in the fluid state throughout.
In another set of 50 runs, we compressed to 
an intermediate pressure, where we expect the crystallization
rate to be higher, and maintained that pressure until we
observed a rapid increase in density. With this
new protocol, the densest lattice is reproduced in seven runs.

We find it remarkable that both our method
and the methods of Refs. \cite{random-perfect,marcotte}
become dramatically less effective at exactly the same dimension. Whereas \textit{a priori},
we might expect that the complexity of the lattice sphere
packing problem rises at a more or less constant exponential
rate as a function of dimension, the evidence of the two cited works and
the present paper raises the hypothesis that the complexity
experiences a sharp increase around $d=20$.
A sharp increase of this kind might indicate a
shift into the glassy regime of the lattice sphere packing
problem. Just as the relaxation time of a fluid increases sharply
as the glass temperature is crossed, it might
be the case here that $d\approx20$  marks the 
beginning of a glassy regime, linked to a sharp increase
in the inverse compression rate required to recover
the densest lattice.

In addition to discovering \textit{de novo} the densest known lattice
in dimension $d$, our method can also be used to discover
suboptimal, yet very dense, extreme lattices. In some dimensions, only a portion
of runs, even at the slowest compression rate attempted,
yielded the densest known lattices (see Table \ref{tabres}). The
identity of the suboptimal lattices produced is
in some cases unexpected, and the second-most-likely-produced
lattice is not always the second-densest known extreme lattice.
For example, in dimension $d=14$, the most frequently produced
lattice after $\Lambda_{14}$ in our simulations is a lattice
(denoted ``dim14kis744'' in Ref. \cite{nebe})
of normalized unit cell volume $2^{14}v=361\sqrt{3}/16$, compared
to $2^{14}v=16\sqrt{3}$ for $\Lambda_{14}$. This is also
the second-densest lattice produced, despite the existence of many extreme lattice
of intermediate density \cite{marcotte}. As was already observed in 
the results of Ref. \cite{marcotte},
we observe a general trend wherein among extreme lattices of equal
density, those with lower kissing numbers (number of neighbors in the
first coordination shell) are more frequently produced,
though this trend is not without exceptions. In the present context,
it makes sense that at finite pressures the basin of an extreme lattice
with a lower kissing number is stabilized by a
greater rattling entropy over the basin of another
extreme lattice of equal density and higher kissing number.

Many of the lattices discovered by our simulations have
not, to our knowledge, been studied before, and we have
submitted them to be archived in the on-line catalog of
lattices \cite{nebe}. A few in particular are definitely worthy of
further study. For example, one of the lattices we
discover in dimension $d=11$ (denoted ``dim11kis422'' in Ref. \cite{nebe}) is equal in density to the
two laminated lattices $\Lambda_{11}^{\text{min,max}}$.
Remarkably, the lattice does
not include any of the laminated lattices $\Lambda_d$ for $d=8,9,10$ as
sublattices of equal minimum norm. This discovery
suggests a possible extension to the conventional lamination
hierarchy described in Refs. \cite{SPLAG,laminated1,laminated2}.

While the focus of the present paper is limited to the
lattice sphere packing problem, the method presented can
easily and naturally be extended to study lattices with
an $n$-element basis for $n>1$. The smallest dimension
in which the densest known packing has a non-trivial
basis is $d=10$, where it has a $40$-element basis \cite{SPLAG}.
In higher dimensions, there are known packings with more
modest basis sizes (for example $n=4,3$ in dimensions $d=20,22$ respectively \cite{ConSlo})
that are denser than the densest known lattices.
Extending our capabilities
to the point of being able to discover \textit{de novo} any
of these nonlattice structures or the densest known
lattices up to $d=24$, would be a major
accomplishment. Of course, discovering yet unknown lattice
and nonlattice packings denser than those constructed analytically
should be considered the ultimate goal.

{\bf Acknowledgements.} I thank {\'E}tienne Marcotte and Salvatore Torquato
for helpful comments.

\bibliography{latt}

\end{document}